\documentclass{elsart}

\usepackage{epsfig}

\usepackage{amssymb}

\begin{document}

\begin{frontmatter}

\title{Synthesis of Mg(B$_{1-x}$C$_x$)$_2$ Powders}

\author{R. H. T. Wilke, S. L. Bud'ko, P. C. Canfield, and D. K. Finnemore}

\address{Ames Laboratory US DOE and Department of Physics and Astronomy, Iowa State University, Ames, IA 50011}

\author{S. T. Hannahs}

\address{National High Magnetic Field Laboratory, Florida State University, 1800 E. Paul Dirac Drive, Tallahassee, Florida 32310}

\begin{abstract}
We have synthesized bulk Mg(B$_{1-x}$C$_x$)$_2$ from a mixture of
elemental Mg, B, and the binary compound B$_4$C. Carbon
incorporation was dramatically improved by a two step reaction
process at an elevated temperature of 1200 $^o$C. This reaction
process results in a solubility limit near x$\sim$0.07. We found
that impurities in the starting B cause an additive suppression of
T$_c$. We combine these data with T$_c$ and H$_{c2}$(T=0) data
from CVD wires as well as plasma spray synthesized powders and
present a unifying H$_{c2}$ and T$_c$ versus x plot.
\end{abstract}
\end{frontmatter}

\section{Introduction}

Superconductivity in MgB$_2$ near 40 K \cite{1} has attracted much
interest due to its potential for applications in the 20-30 K
range. The low upper critical field \cite{2} and high anisotropy
ratio, $\gamma$=H$^{\parallel ab}_{c2}$/H$^{\perp ab}_{c2}$
\cite{3,4,5}, of pure MgB$_2$, limit its potential usefulness
\cite{6}. Carbon doping of MgB$_2$ has been shown to be an
effective method for enhancing H$_{c2}$ \cite{7,8,9} while
simultaneously decreasing the anisotropy ratio by increasing
H$^{\perp ab}_{c2}$ more rapidly than H$^{\parallel ab}_{c2}$
\cite{10,11}. Carbon doped MgB$_2$ filaments require high reaction
temperatures which lead to large grain sizes and poor J$_c$
values, in spite of the enhancements in H$_{c2}$ \cite{16}.
Reports of successful fabrication of superconducting wire using
powder-in-tube processing \cite{12,13}, coupled with the fact that
J$_c$ values in powder samples can readily be increased by the
addition of various particles \cite{30,31,32,33} motivate a desire
for synthesizing single phase carbon doped MgB$_2$ powder. Carbon
doped bulk polycrystalline MgB$_2$ with approximately 10\% carbon
incorporation has previously been synthesized by mixing elemental
Mg and the binary B$_4$C \cite{14,15}. Systematic carbon doping of
Mg(B$_{1-x}$C$_x$)$_2$ with x$<$0.10 has been achieved in single
crystals \cite{8,23} and polycrystalline wires fabricated by
chemical vapor deposition (CVD) \cite{7,16}. In this paper we
explore the possibility of preparing Mg(B$_{1-x}$C$_x$)$_2$ with
x$<$0.10 using B$_4$C as the carbon source.

\section{Experimental Methods}

Powder samples of Mg(B$_{1-x}$C$_x$)$_2$ were prepared in a two
step process. First, stoichiometric mixtures of distilled Mg,
elemental B, and the binary compound B$_4$C were reacted for 48
hours at 1200 $^o$C. The resultant sample was then reground in
acetone, pressed into a pellet, and re-sintered for an additional
48 hours at 1200 $^o$C. Two different batches were made. The first
with 0.995 purity B (metals basis) from Alpha Aesar, and the
second with 0.9997 purity isotopically enriched $^{11}$B from
Eagle Picher. The three main impurities in the 0.995 purity B are
C, Si, and Fe, which have relative atomic abundances of 0.25\% and
0.20\%, and 0.10\% respectively. The isotopically enriched
$^{11}$B contained 0.02\% Ta, 0.001\% Cu, and 0.001\% Fe. Samples
made with isotopically enriched $^{11}$B were shown to have a
higher residual resistivity ratio (RRR) than samples made with
nominal 0.9999 purity B, indicating the isotopic enrichment
process yields perhaps the purest boron available \cite{19}.
B$_4$C from Alpha Aesar was used as the carbon source in both
runs. The B$_4$C had a nominal purity of 0.994 metals basis with
the two primary impurities being Si and Fe, which occur in
relative abundances of 0.37\% and 0.074\% respectively, values
similar to those in the 0.995 purity B. Since impurities in the
boron have been shown to suppress T$_c$ \cite{19}, two differing
boron purities were used to examine the effects of the starting
boron purity on the T$_c$ and H$_{c2}$(T=0) values in carbon doped
MgB$_2$. We found it necessary to use multiple reaction steps to
incorporate the carbon as uniformly as possible. To avoid
confusion regarding the meaning of x in Mg(B$_{1-x}$C$_x$)$_2$, we
will henceforth refer to the nominal carbon content as x$_n$ and
the inferred carbon content after the m-th reaction step as
x$_{im}$.

Powder x-ray diffraction (XRD) measurements were made at room
temperature using CuK$\alpha$ radiation in a Rigaku Miniflex
Diffractometer. A silicon standard was used to calibrate each
pattern. Lattice parameters were determined from the position of
the MgB$_2$ (002) and (110) peaks. DC magnetization measurements
were performed in a Quantum Design MPMS-5 SQUID magnetometer.
Transport measurements were done using a four probe technique,
with platinum wires attached to the samples with Epotek H20E
silver epoxy. Resistance versus temperature in applied fields up
to 14 T were carried out in a Quantum Design PPMS-14 system and
resistance versus field was measured up to 32.5 T using a lock-in
amplifier technique in a resistive DC magnet at the National High
Magnetic Field Laboratory in Tallahassee, Florida.

\section{Results and Discussion}

Using the lower purity boron, we reacted a sample with nominal
carbon content of x$_n$=0.05 for 48 hours at 1200 $^o$C. The
carbon content can be estimated by the shift of the x-ray (110)
peak position relative to that of a nominally pure MgB$_2$ sample
made under the same conditions \cite{7,15}. Although the B$_4$C is
a different B source than that used for the reference sample, the
0.994 purity B$_4$C and 0.995 purity B contain similar
concentrations of impurity phases, and we cautiously proceed with
estimates of the carbon content ignoring minor differences between
boron sources. Indexing of the (110) peak for the pure sample and
that containing a nominal carbon content of x$_n$=0.05 yielded an
inferred carbon level after this first reaction step of
approximately x$_{i1}$=0.031. In order to ensure the carbon was
fully incorporated and uniformly distributed within this sample,
it was reground in acetone, pressed into a pellet, and sintered
for an addition 48 hours at 1200 $^o$C. The subsequent sample
showed a further increase in the (110) peak position (Figure 1)
which yielded an inferred carbon content of x$_{i2}$=0.069. This
sample also showed decrease in T$_c$ (Figure 2), which is
consistent with more carbon being incorporated in the structure.
In addition to shifting to higher 2$\theta$, the (110) peak became
sharper. The full-width-at-half-maximum (FWHM) decreased from
0.221$^o$ to 0.157$^o$ after the second sintering step. Therefore
the second sintering step not only incorporated more carbon but it
appears to have resulted in a more uniform carbon distribution. It
should be noted that after the two sintering steps the FWHM values
of the MgB$_2$ peaks were comparable to those of the Si standard,
indicating we have achieved a fairly high level of homogeneity. It
is also worth noting that whereas the (110) peak shifts and
sharpens as a result of a second reaction step, the (002) peak
does neither, having FWHM values comparable to, but slightly
larger than, the neighboring Si (311) peak after both reaction
steps. This indicates that the c-axis spacing and periodicity are
particularly insensitive to this degree of carbon doping and/or
disorder.

Whereas the x$_{i2}$ value exceeds the nominal value of x$_n$=0.05
in the starting material, the presence of MgB$_4$, as evidenced by
strong peaks in the x-ray spectrum (Figure 3) may account for the
discrepancy if we assume no carbon enters the MgB$_4$ structure.
Comparison of the x-ray spectra for the single and two step
reactions shows an increase in the intensity of the MgB$_4$ peaks
after the second reaction step. This step was done without the
addition of any extra Mg to compensate for potential Mg loss. It
is possible that while sintering at 1200 $^o$C, some Mg is driven
out of the MgB$_2$ structure and this loss results in conversion
of MgB$_2$ to MgB$_4$, with the excess Mg forming MgO and possibly
condensing on the walls of the tantalum reaction vessel during the
quench. To determine whether or not Mg loss is responsible for the
apparent increase in the carbon content, the second sintering step
for a sample with nominal concentration x$_n$=0.05 was carried out
in an atmosphere of excess Mg vapor. This sample exhibited a T$_c$
of 29.8 K and a shift in the (110) x-ray peak yielding an inferred
carbon concentration of x$_{i2}$=0.050 (Figure 4), consistent with
the nominal concentration. Thus the apparent difference in carbon
content for samples which undergo a second sintering step without
the presence of excess Mg to compensate for Mg loss relative to
those which undergo the second sintering step with excess Mg is
presumably the result of a fixed amount of carbon being
incorporated into a decreased amount of MgB$_2$. To avoid the
potential creation of percolation networks of Mg within the
samples we chose to perform the second sintering step without any
additional Mg.

Using 0.995 purity boron and 0.994 purity B$_4$C, an entire series
with nominal carbon levels of x$_n$=0, 0.0125, 0.025, 0.035, 0.05,
and 0.075 was prepared using the two step reaction profile. The
(002) and (110) x-ray peak positions for the entire series are
plotted in figure 5. The (002) peak position is roughly constant
for all carbon levels, consistent with the results found by Wilke
et al. \cite{7} and Avdeev et al. \cite{15}, which showed only a
slight expansion along the c-axis for carbon doping levels up to
10$\pm$2\%. The (110) peak position shifts towards higher
2$\theta$ values as x is increased up to x$_n$=0.05, at which
point it appears to be saturated. Using the (110) peak position
for the nominally pure sample as our standard, the inferred carbon
concentrations for the entire series are x$_{i2}$=0, 0.01, 0.034,
0.044, 0.069, and 0.067. The samples with carbon concentrations
saturating near x$_{i2}$$\sim$0.07 show an increase in the
MgB$_2$C$_2$ phase as a function of the nominal carbon content
(Figure 5b). Thus the excess carbon is precipitating out as
MgB$_2$C$_2$.

Normalized magnetization measurements (Figure 6) confirm the
highest two doping levels have incorporated roughly the same
amount of carbon. Their transition temperatures all lie slightly
below 28 K. Defining T$_c$ using a 2\% screening criteria the
x$_{i2}$=0.069 and 0.067 levels have T$_c$ values of 27.5 K and
27.8 K respectively. For these higher doping levels, the nominal
concentrations did not yield systematic increases in carbon level,
but the change in the a-lattice parameter and T$_c$ are consistent
with one another; i.e. samples which apparently incorporated more
carbon had smaller a-lattice parameters and lower T$_c$ values.

This saturation near x$_{i2}$$\sim$0.07 is not entirely unexpected
given the results of 10$\pm$2\% carbon incorporation using B$_4$C
reported by Ribeiro et al. \cite{14} and Avdeev et al. \cite{15}.
In optimizing the reaction, Ribeiro found that under certain
conditions, T$_c$ values below the near 22 K reported for the
optimal 24 hours at 1100 $^o$C reaction could be attained,
suggesting higher carbon content phases may be metastable. To test
whether the saturation we observed was an effect due to the use of
a two step reaction, as opposed to the single step employed by
Ribeiro et al., we repeated their work, making a sample of nominal
concentration Mg(B$_{0.8}$C$_{0.2}$)$_2$ using only Mg plus
B$_4$C. This sample underwent an initial 48 hour reaction at 1200
$^o$C to form the superconducting phase and a second sintering for
48 hours at 1200 $^o$C. After the first 48 hours at 1200 $^o$C we
find a superconducting phase with T$_c$ near 22 K, and a lattice
parameter shift which yielded inferred values of x$_{i1}$=0.092
slightly less than the x$_i$=0.10 obtained by Ribeiro and
coworkers using isotopically enriched $^{11}$B$_4$C as the carbon
source \cite{15}. After the second sintering step, T$_c$ rises to
27.9 K. The (110) x-ray peaks shifted to lower 2$\theta$, yielding
an inferred x$_{i2}$=0.065 (Figure 7). Although two reaction steps
were used, no change in the FWHM of the (110) peak was observed.
As in the case of the x$_n$=0.075 sample, the decrease in carbon
content could be due to carbon precipitating out in the form of
MgB$_2$C$_2$. The relative intensity of the most prominent
MgB$_2$C$_2$ peak to that of MgB$_2$ approximately doubles going
from the single step reaction to the two step reaction. In order
to check if more carbon would be precipitated out in the form of
MgB$_2$C$_2$ by simply adding more sintering steps to the growth
process an additional sample underwent a three step reaction: an
initial 48 hours at 1200 $^o$C to form the superconducting phase
followed by two additional sintering steps of 48 hours at 1200
$^o$C. After this third reaction step, the sample exhibited a
T$_c$ of 27.9 K and the (110) peak position yielded an inferred
carbon content of x$_{i3}$=0.064 (Figure 7). These values are
comparable to our previous results with only two sintering steps.
Thus the carbon content appears to saturate in the vicinity of an
inferred carbon content of x$_i$=0.065. The fact that saturation
near x$_{i2}$=0.065 occurred for samples which had x$_{i1}$ both
above and below this level indicates that in equilibrium the
solubility limit for 1200 $^o$C reactions near 1 atm is in the
range 0.065$<$ x$_i$ $<$ 0.07.

Transport measurements were made in order to determine the upper
critical field. An onset criteria was used in both resistance
versus temperature and resistance versus field. Figure 8a plots
resistance versus temperature in applied fields up to 14 T and
figure 8b plots resistance versus field at temperatures down to
1.4 K for the sample with a carbon level of x$_{i2}$=0.069
(x$_n$=0.05). The zero field resistive transition has a width of
less than 1 K, but this significantly broadens as the strength of
the applied field increases. The R vs. H measurements show a
related broadening. For example, the width of the transition at
1.4 K is nearly 20 T wide. This should be compared to the 10 T
wide, approximately linear transitions reported for 5.2\% carbon
doped filaments \cite{16}. Optical images taken under polarized
light show the superconducting grains in the powder sample are
5-10 $\mu$m in size. In the case of the wire sample a majority of
the grains are in the 1-5 $\mu$m range \cite{16}. We therefore
ascribe the increased width of this transition to a combination of
poor flux pinning due to the large grain size associated with the
high reaction temperature as well as to possible remaining
inhomogeneities in carbon incorporation within the sample.

H$_{c2}$ curves for x$_{i2}$=0.034 and x$_{i2}$=0.069 along with a
pure wire \cite{2} and a carbon doped wire with an inferred carbon
content of x$_i$=0.052 \cite{7} are plotted in figure 9. The
powder sample with x$_{i2}$=0.034 has a T$_c$ slightly less than
that of the carbon doped wire with x$_i$=0.052 and an
H$_{c2}$(T=0) more than 5 T lower. This marked difference shows
that for carbon doped samples made with differing nominal boron
purities, T$_c$ alone is not a good caliper of H$_{c2}$(T=0). The
sample with inferred carbon content of x$_{i2}$=0.069 has a T$_c$
nearly 7 K below the aforementioned wire and an H$_{c2}$(T=0) just
above 30 T. Comparing the two powder samples to one another, we
see an increase in the slope of H$_{c2}$ near T$_c$ for the higher
doping level, which results in a higher H$_{c2}$(T=0), consistent
with our earlier findings \cite{7,16}.

Carbon has been shown to suppress T$_c$ at a rate of roughly 1
K/\%C for up to 5\% carbon substitution \cite{16}. The
magnetization and transport measurements indicate T$_c$ of the
powder samples made with the 0.995 purity Alpha Aesar boron is
also being suppressed at a rate of roughly 1 K/\%C, but relative
to the suppressed, near 37 K, transition temperature of the
nominally pure sample. The suppressed transition temperature of
the nominally pure MgB$_2$ sample lies approximately 2 K below
results obtained using high purity natural boron wires \cite{77}.
MgB$_2$ made from lower purity boron has been shown to have lower
transition temperatures \cite{19}. In figure 10, a comparison of
T$_c$ versus $\mid$$\Delta$a$\mid$ for carbon doped samples
prepared with lower purity boron to carbon doped wires made with
high purity boron shows the manifold associated with the impure
boron powder is shifted downward by approximately 2 K for all
carbon levels. To confirm that this difference is a result of the
purity of the starting boron, a second set of two step process
samples made with isotopically enriched $^{11}$B were measured.
The results are included in figure 10. Also included is a set of
carbon doped powders made by a plasma spray process \cite{20}. The
agreement between the CVD wires, plasma spray powders, and
$^{11}$B samples shows that high purity boron in a variety of
forms responds to carbon doping in a similar manner. These data
also seem to indicate that there is some additional impurity in
the 0.995 pure Alpha Aesar boron that is systematically
suppressing T$_c$.

Figure 11 plots a comparison of the (002) and (110) x-ray peaks of
pure MgB$_2$ using the three different purity levels of boron. The
sample made with the nominal 0.995 purity boron shows a shift of
the (110) peak to higher 2$\theta$ by 0.09$^o$, which, if it were
to be associated with carbon doping, would be consistent with
carbon doping of approximately 1.8\%. This level far exceeds the
stated carbon impurity level of 0.25\% in the 0.995 purity B as
claimed in the certificate of analysis provided by Alpha Aesar. To
check whether by using lower purity boron we have inadvertently
doped with carbon to such a high level, we measured the resistive
onset of superconductivity in an externally applied 14 T field for
the nominal pure MgB$_2$ using the 0.995 purity boron and compared
the temperature with those attained for carbon doped fibers
reported in reference \cite{16}. MgB$_2$ fibers reacted at 1200
$^o$C for 48 hours showed an onset of superconductivity in an
externally applied magnetic field of 14 T at 10.2 K, 14.8 K, and
18.5 K for pure, 0.6\%, and 2.1\% carbon doping \cite{16}. If the
shift of the (110) peak in the nominally pure MgB$_2$ made from
0.995 purity boron were a result of inadvertent carbon doping, we
would expect an onset of superconductivity in an applied 14 T
field at a temperature between 15 K and 18 K. However, such a
measurement yielded an onset near 13K indicating if carbon is
present as an impurity, it is less than 0.6\%, which consistent
with the estimate provided by Alpha Aesar. Therefore the manifold
of T$_c$ versus $\mid$$\Delta$a$\mid$ for the lower purity boron
is shifted downward by some as of yet unidentified impurity
associated with the Alpha Aesar boron.

H$_{c2}$ values were determined using an onset criteria in
resistivity versus temperature and resistance versus field
measurements. Pellets made using the isotopically enriched
$^{11}$B lacked structural integrity and were unsuitable for
transport measurements. Therefore we could only attain
H$_{c2}$(T=0) values only for samples made from the CVD wires,
plasma spray powders, and the 0.995 purity boron (Figure 12). For
the dirtier, 0.995 purity powder, at a doping level of
x$_{i2}$=0.034 the upper critical field agrees with the results of
a carbon doped wire with an inferred carbon content of x$_i$=0.038
from reference \cite{7}. At doping levels near x$_{i2}$=0.065,
H$_{c2}$ values fall several Tesla below the manifold for "clean"
carbon doped samples. In carbon doped MgB$_2$, enhancement of
H$_{c2}$ due to scattering effects \cite{11,22} competes with the
suppression of T$_c$ caused by electron doping \cite{11}. By
further suppressing T$_c$ by introducing additional impurities in
the system, we may have limited the maximum H$_{c2}$ attainable
through carbon doping.

\section{Conclusions}

We have established a method for synthesizing
Mg(B$_{1-x}$C$_x$)$_2$ using a mixture of distilled magnesium,
boron and the binary compound B$_4$C. Impurities in the starting
boron effect T$_c$ and the magnitude of the a-lattice parameter.
By tracking $\mid$$\Delta$a$\mid$ and T$_c$ we were able to show
that different boron purities lead to differing
T$_c$($\mid$$\Delta$a$\mid$) manifolds. There appears to be a
solubility limit in the carbon content for samples synthesized at
1200 $^o$C and 1 atm near x$\sim$0.07. Lower purity boron in the
starting material results in lower transition temperatures and
appears to limit the maximum achievable upper critical field.

\clearpage

\begin{figure}
\begin{center}
\includegraphics[angle=0,width=180mm]{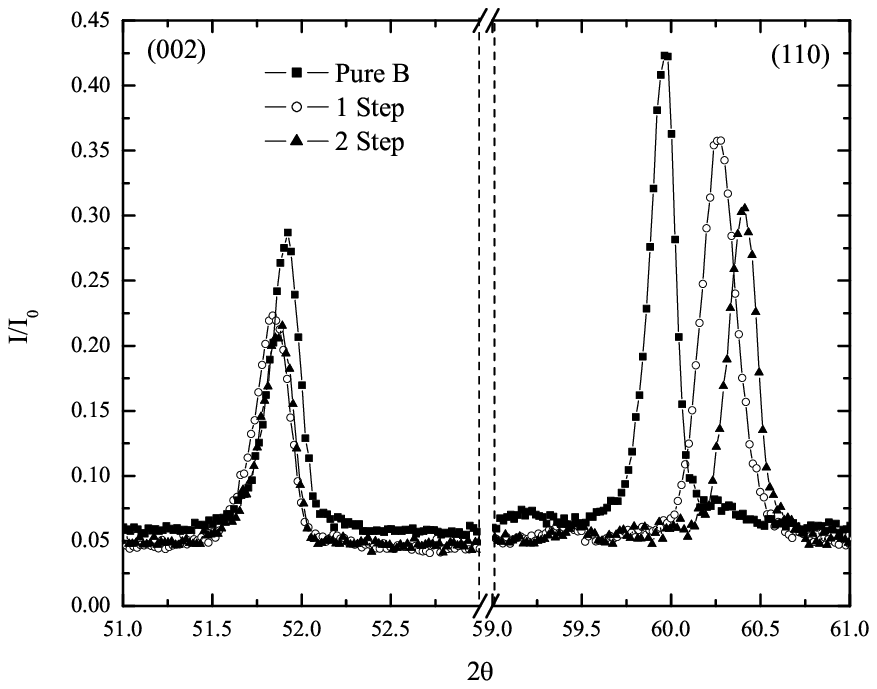}
\end{center}
\caption{Powder x-ray diffraction (002) and (110) peaks for pure
and nominal Mg(B$_{.95}$C$_{.05}$)$_2$ samples made using the
0.995 purity boron as the starting material. The pure sample was
reacted using two steps. For the carbon doped sample, the second
sintering step shifts the (110) peak position to higher 2$\theta$
indicating the incorporation of a higher carbon
concentration.}\label{f1}
\end{figure}

\clearpage

\begin{figure}
\begin{center}
\includegraphics[angle=0,width=180mm]{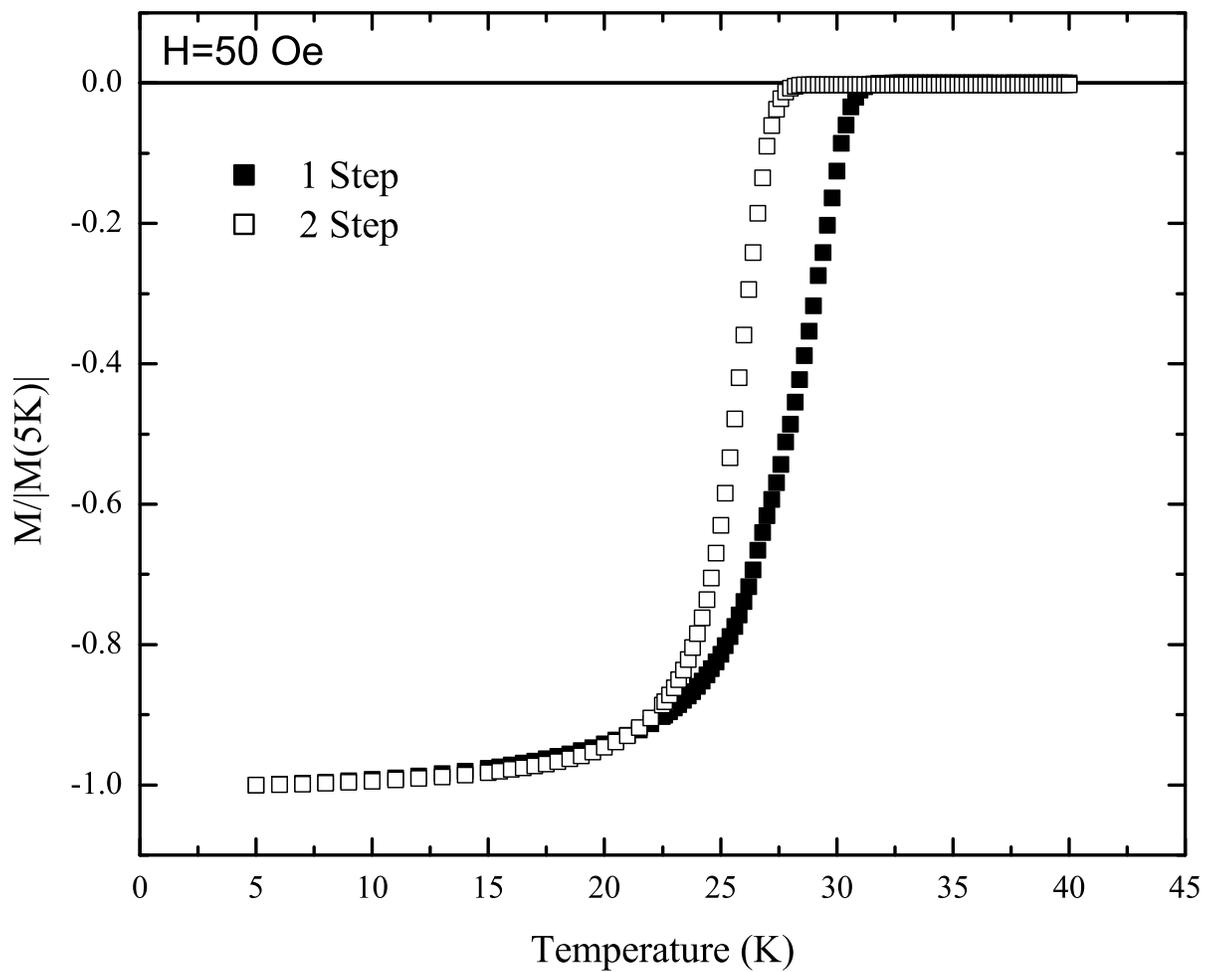}
\end{center}
\caption{Normalized magnetization curves for nominal
Mg(B$_{.95}$C$_{.05}$)$_2$ samples. The second sintering step
lowers T$_c$, consistent with the incorporation of a higher carbon
concentration.}\label{f2}
\end{figure}

\clearpage

\begin{figure}
\begin{center}
\includegraphics[angle=0,width=180mm]{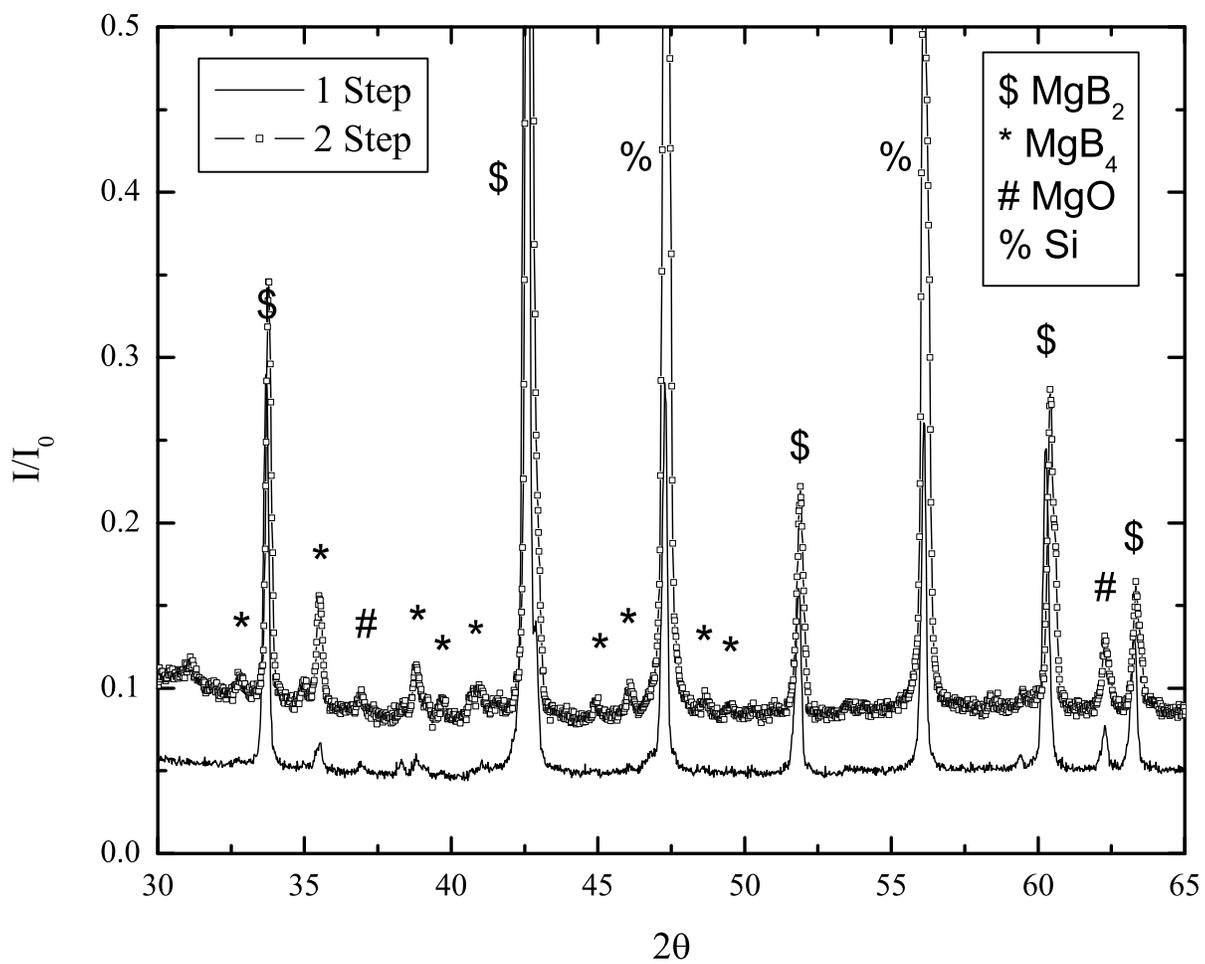}
\end{center}
\caption{Normalized powder x-ray pattern for a sample with
x$_n$=0.05 synthesized using a one step and a two step reaction.
The two step sample clearly contains enhanced amounts of
MgB$_4$.}\label{f3}
\end{figure}

\clearpage

\begin{figure}
\begin{center}
\includegraphics[angle=0,width=180mm]{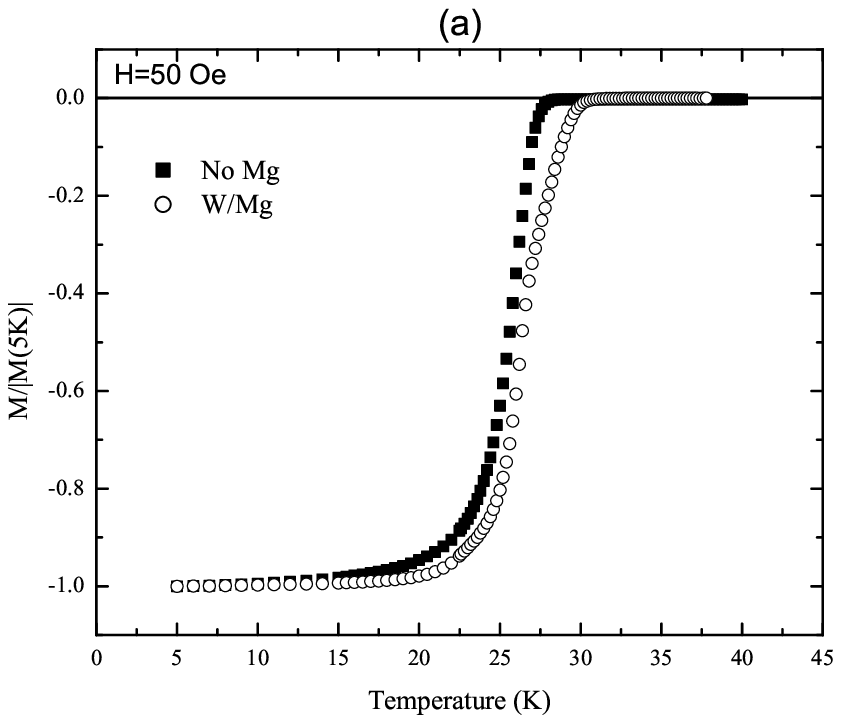}
\end{center}
\end{figure}

\clearpage

\begin{figure}
\begin{center}
\includegraphics[angle=0,width=180mm]{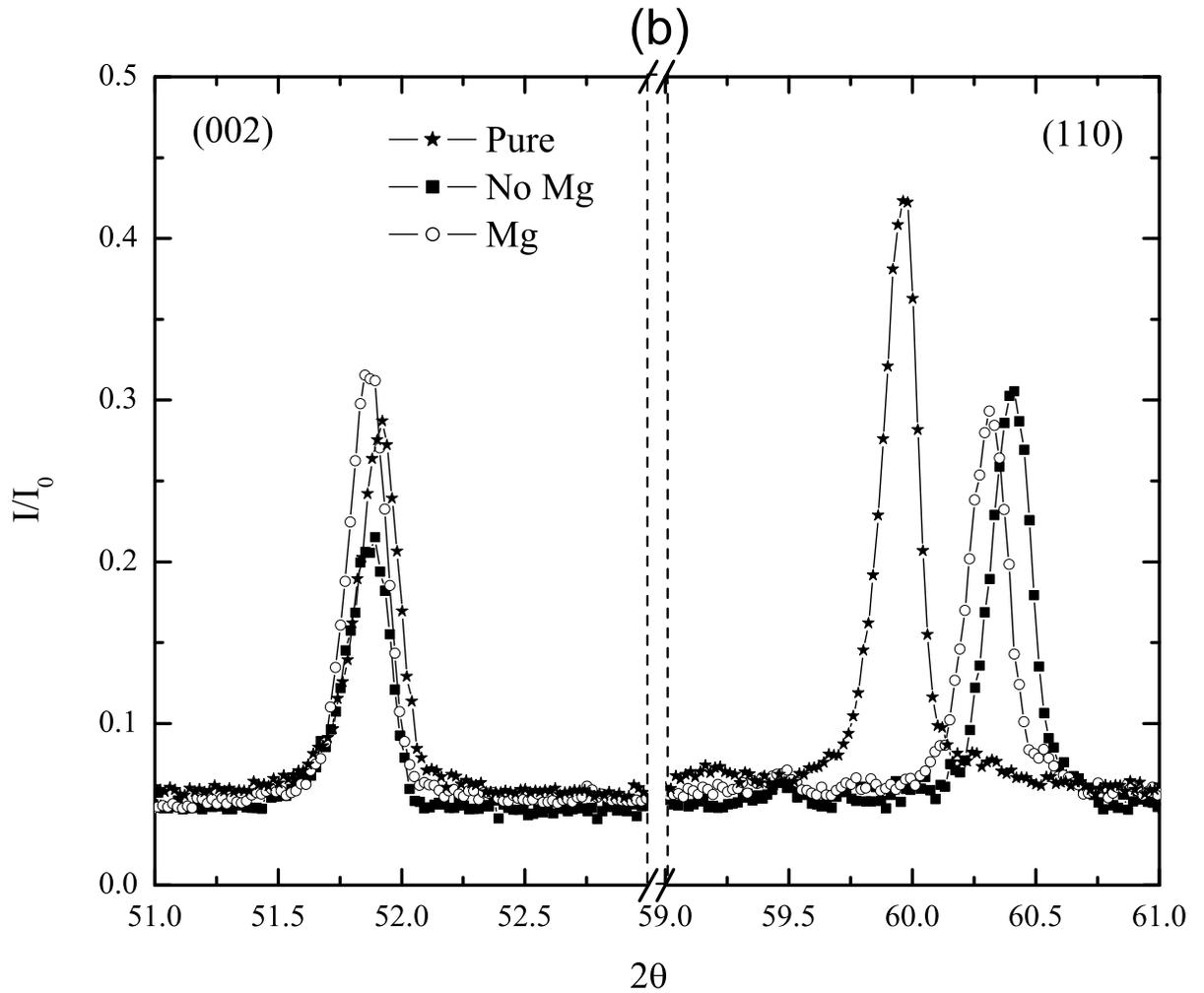}
\end{center}
\caption{(a) Normalized magnetization curves and (b) x-ray (002)
and (110) peaks for a sample of x$_n$=0.05 reacted using a two
step process. If the second sintering step is performed without
any excess Mg to compensate for potential losses, the resultant
carbon content within the MgB$_2$ phase is increased.}\label{f4}
\end{figure}

\clearpage

\begin{figure}
\begin{center}
\includegraphics[angle=0,width=180mm]{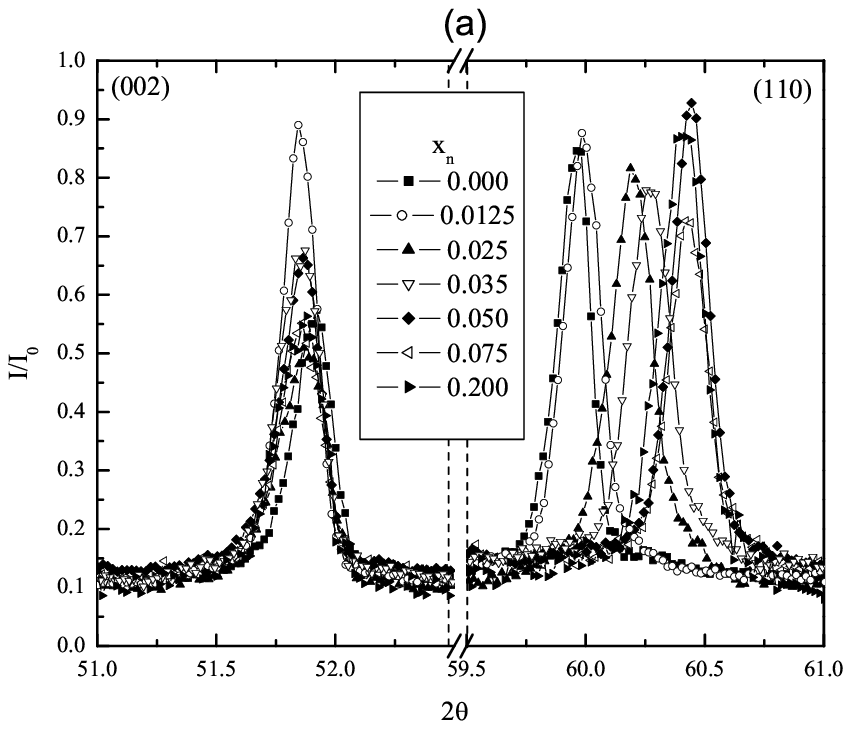}
\end{center}
\end{figure}

\clearpage

\begin{figure}
\begin{center}
\includegraphics[angle=0,width=180mm]{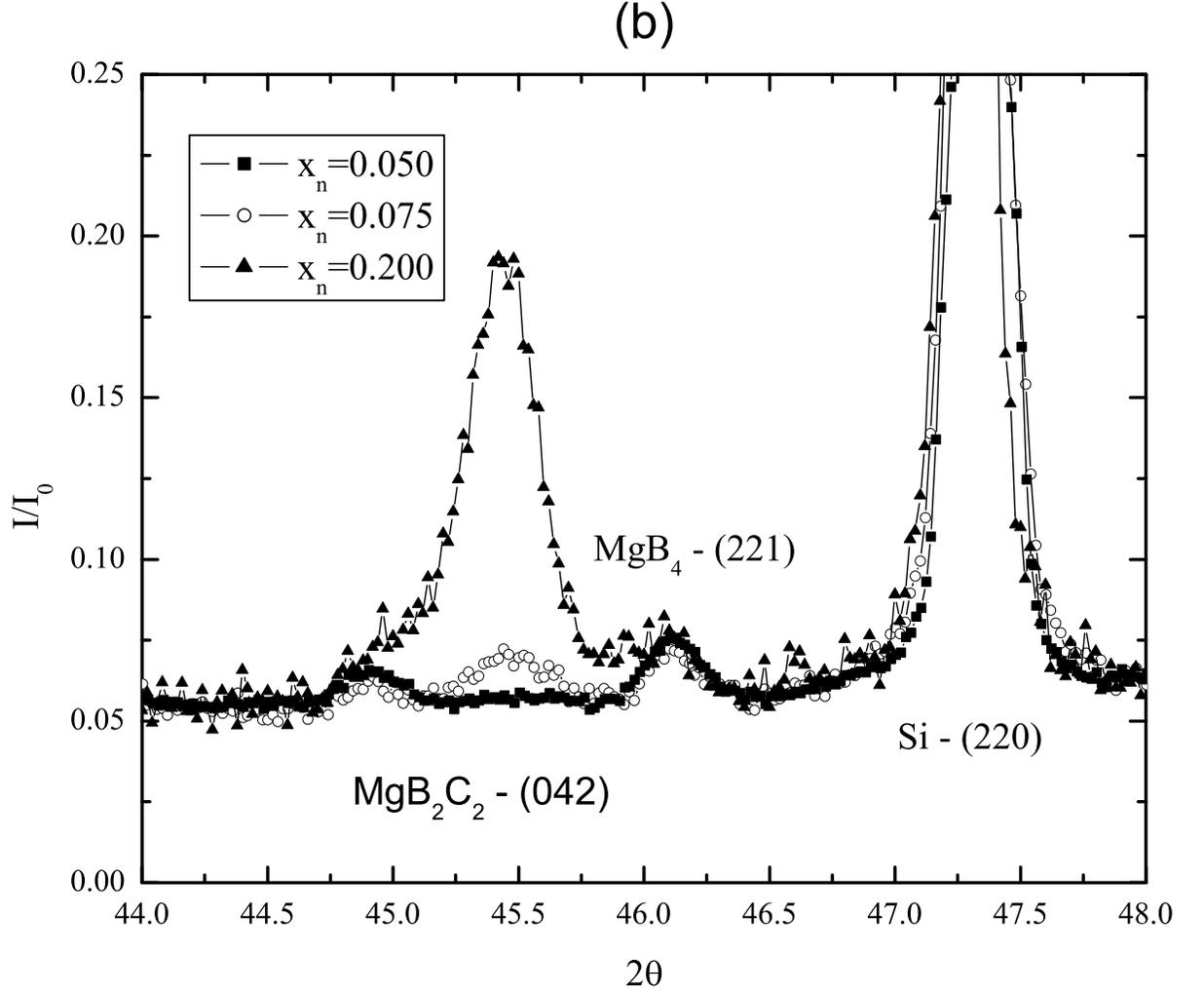}
\end{center}
\caption{(a) Evolution of the (002) and (110) x-ray peaks for
Mg(B$_{1-x}$C$_{x}$)$_2$ samples with nominal x$_n$=0, 0.0125,
0.025, 0.035, 0.05, 0.075, and 0.20 synthesized using a two step
reaction. The shift of the (110) peak relative to that of the
un-doped yields inferred carbon concentrations of x$_{i2}$=0.01,
0.034, 0.044, 0.069, 0.067, 0.065. (b) For samples saturating near
x$_{i2}$=0.07 the excess carbon precipitates out in the form of
MgB$_2$C$_2$ as can be seen by the emergence of the MgB$_2$C$_2$
(042) peak as a function of nominal carbon content.}\label{f5}
\end{figure}

\clearpage

\begin{figure}
\begin{center}
\includegraphics[angle=0,width=180mm]{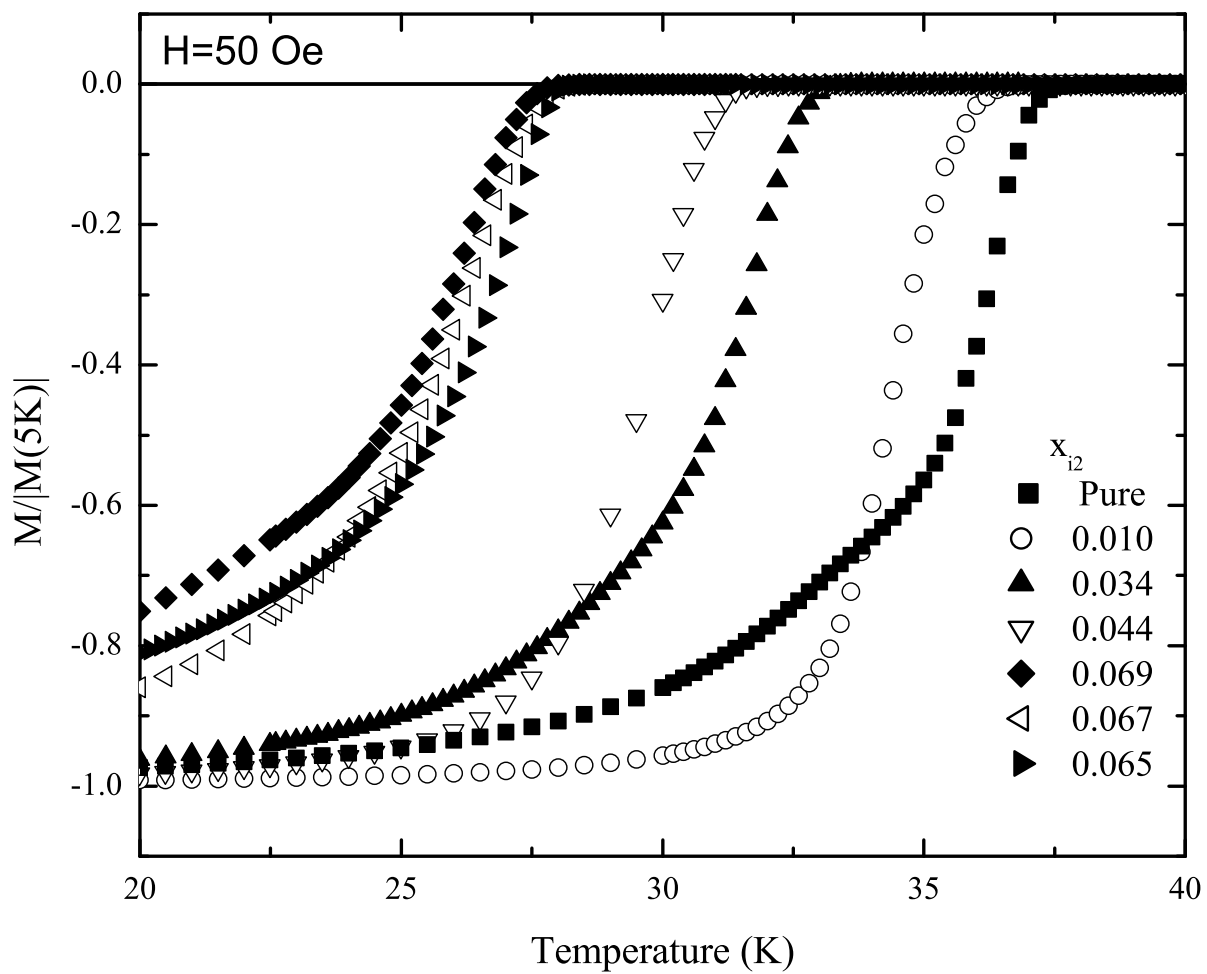}
\end{center}
\caption{Normalized magnetic transitions for the series of
Mg(B$_{1-x}$C$_{x}$)$_2$ with x$_{i2}$=0.01, 0.034, 0.044, 0.069,
0.067, 0.065, synthesized with 0.995 purity B and reacted using a
two step process.}\label{f6}
\end{figure}

\clearpage

\begin{figure}
\begin{center}
\includegraphics[angle=0,width=180mm]{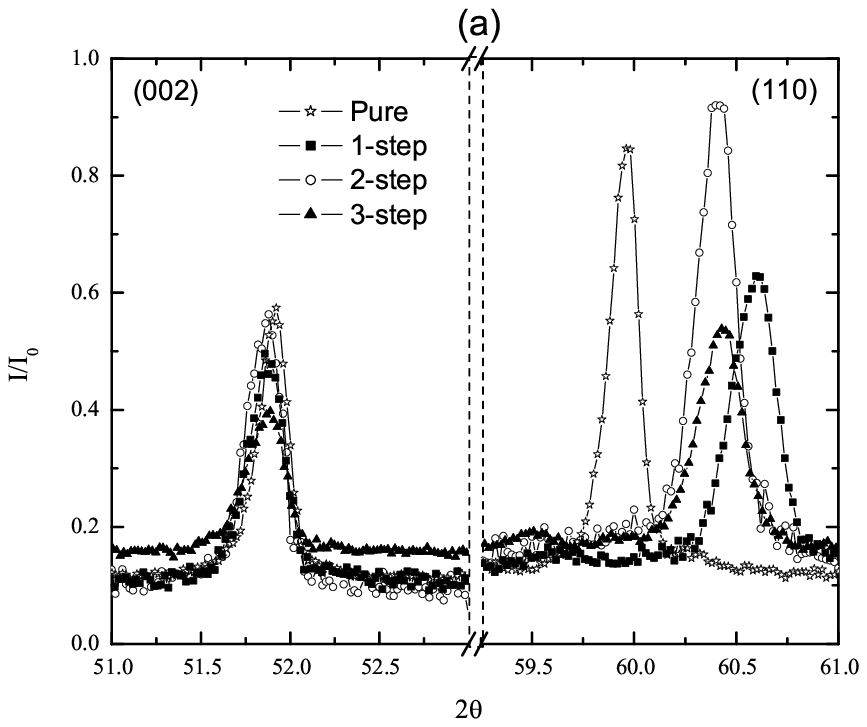}
\end{center}
\end{figure}

\clearpage

\begin{figure}
\begin{center}
\includegraphics[angle=0,width=180mm]{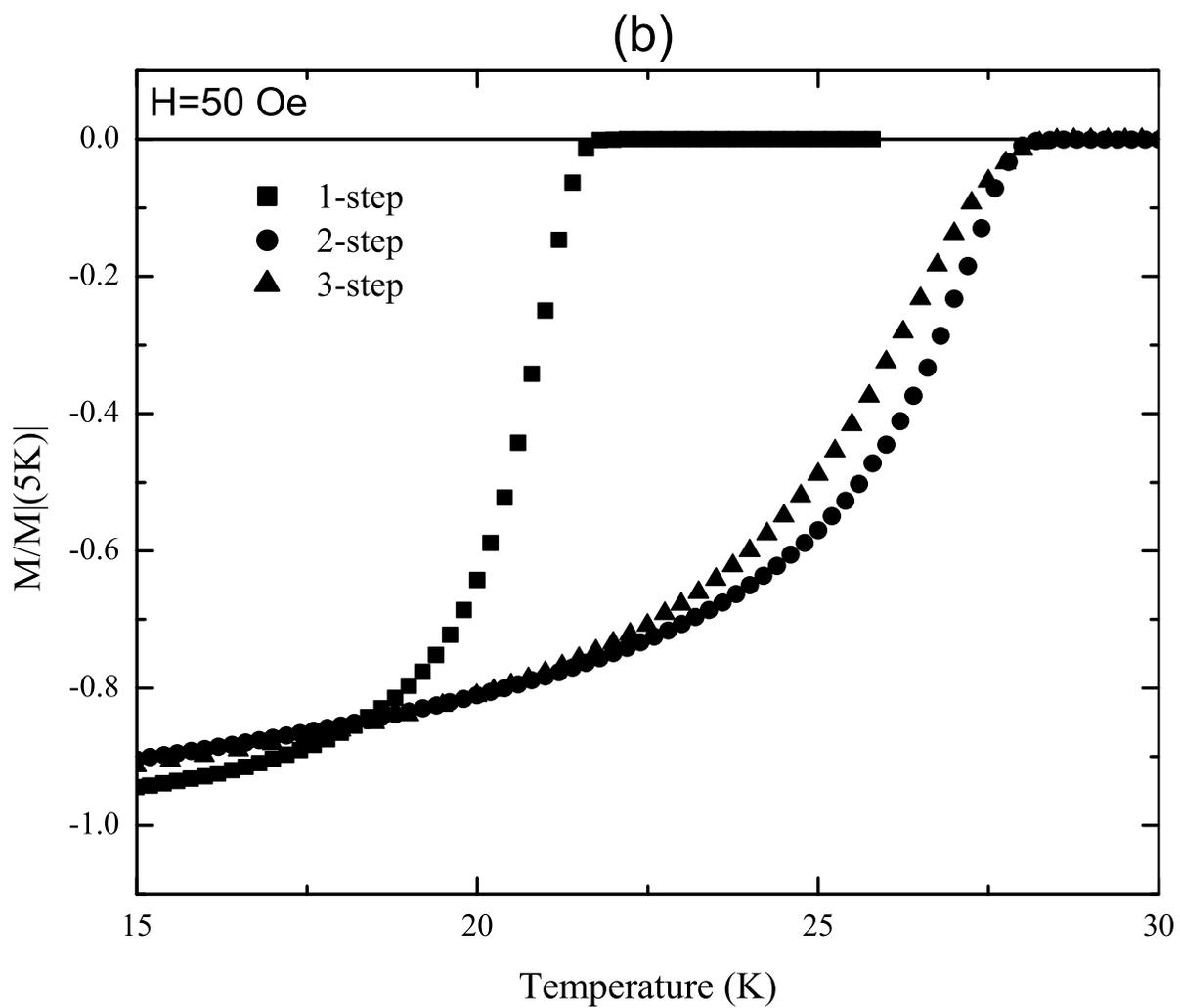}
\end{center}
\caption{(a) (002) and (110) x-ray peaks for nominal
Mg(B$_{0.8}$C$_{0.2}$)$_2$ using B$_4$C as the boron and carbon
source and reacted using 1, 2, and 3 step reaction processes. (b)
Normalized magnetic transitions for these samples. }\label{f7}
\end{figure}

\clearpage

\begin{figure}
\begin{center}
\includegraphics[angle=0,width=180mm]{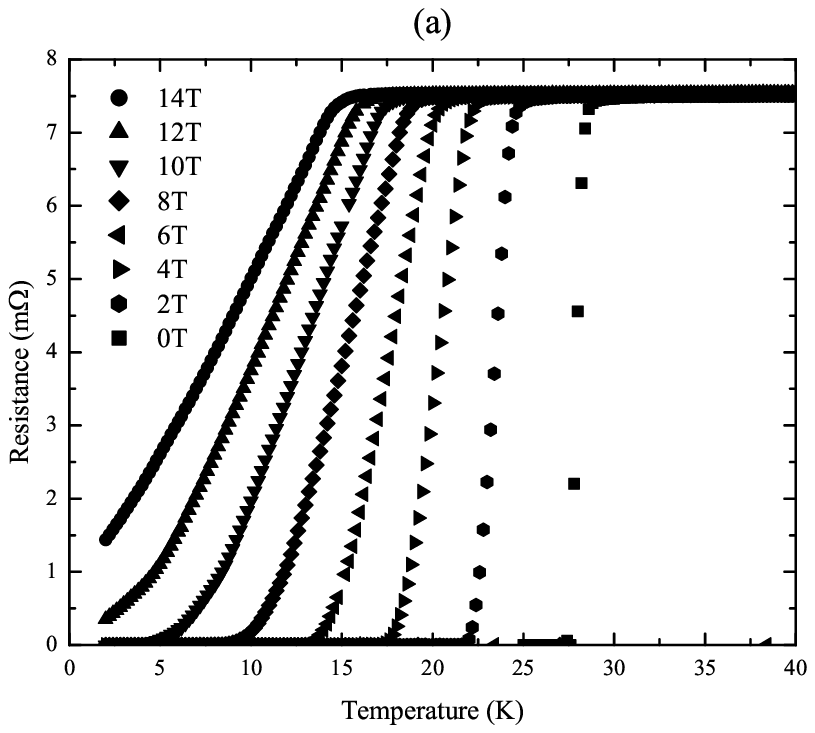}
\end{center}
\end{figure}

\clearpage

\begin{figure}
\begin{center}
\includegraphics[angle=0,width=180mm]{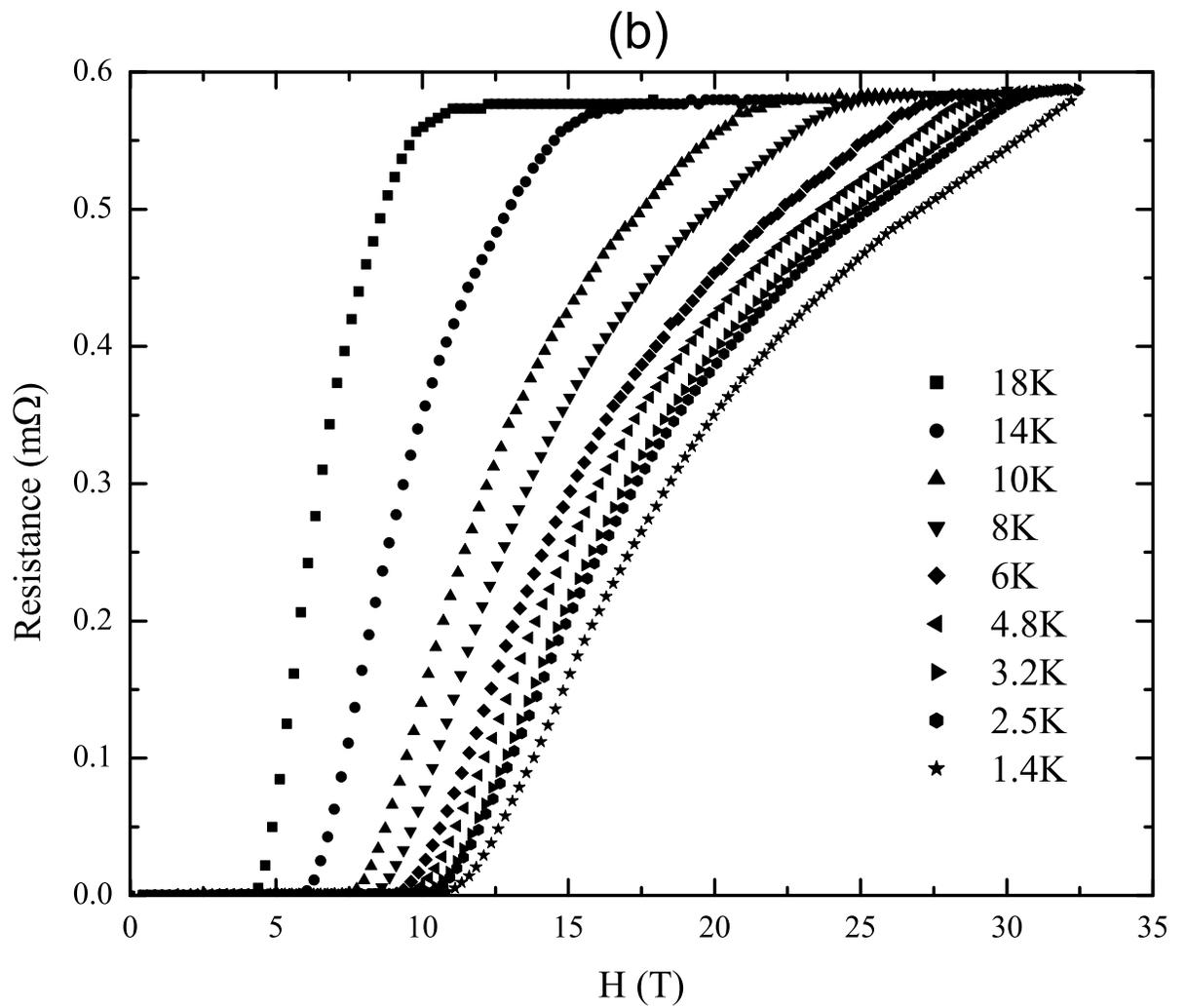}
\end{center}
\caption{(a) Resistance versus temperature and (b) resistance
versus field for a sample with x$_{i2}$=0.069. }\label{f8}
\end{figure}

\clearpage

\begin{figure}
\begin{center}
\includegraphics[angle=0,width=180mm]{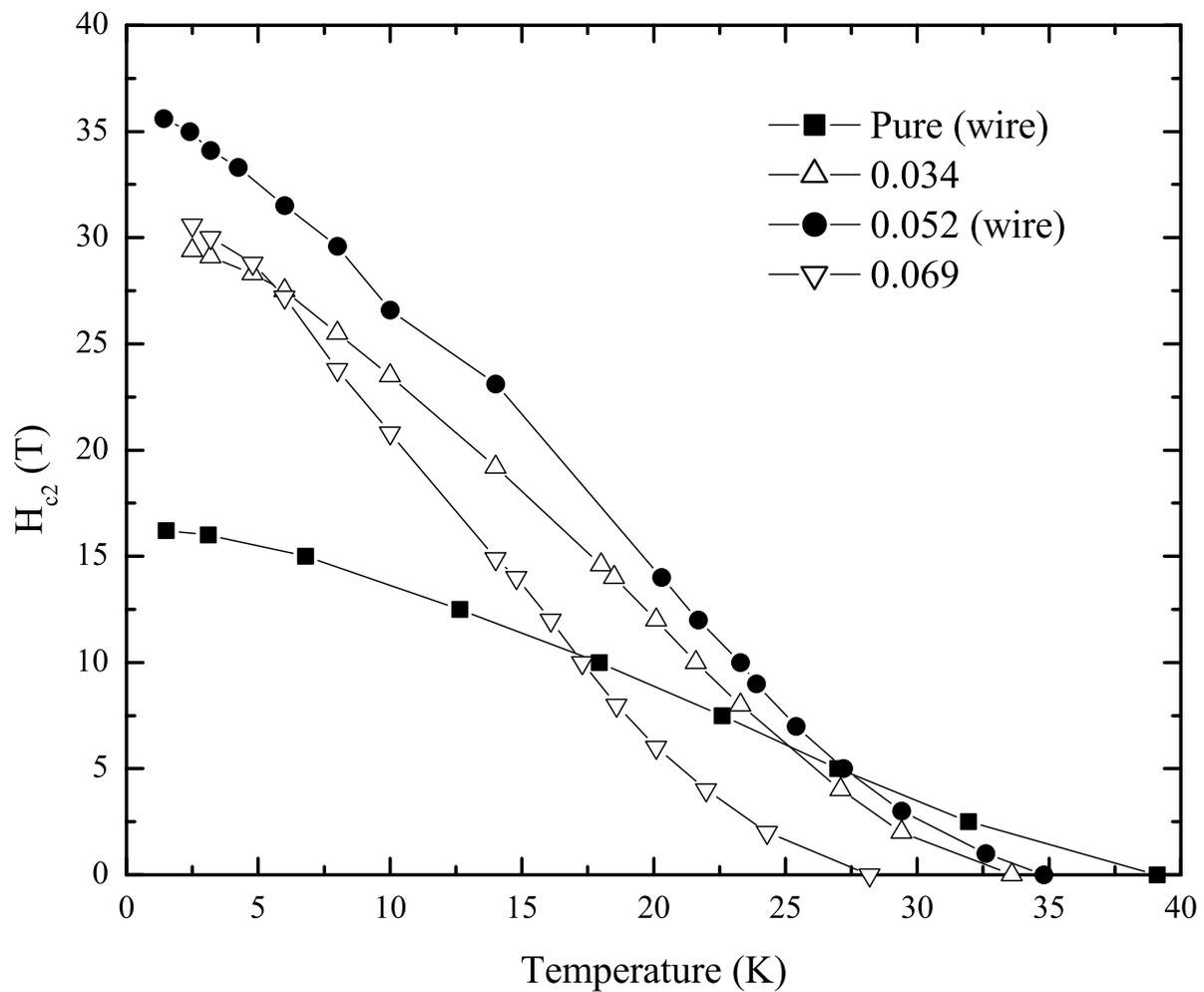}
\end{center}
\caption{Comparison of upper critical field curves for pellets
with carbon doping levels of x$_{i2}$=0.034 and 0.069 with wires
containing x$_i$=0 and 0.052. The x$_i$=0 and 0.052 samples were
made by reacting Mg vapor with boron filaments, see reference
\cite{7}.}\label{f9}
\end{figure}

\clearpage

\begin{figure}
\begin{center}
\includegraphics[angle=0,width=180mm]{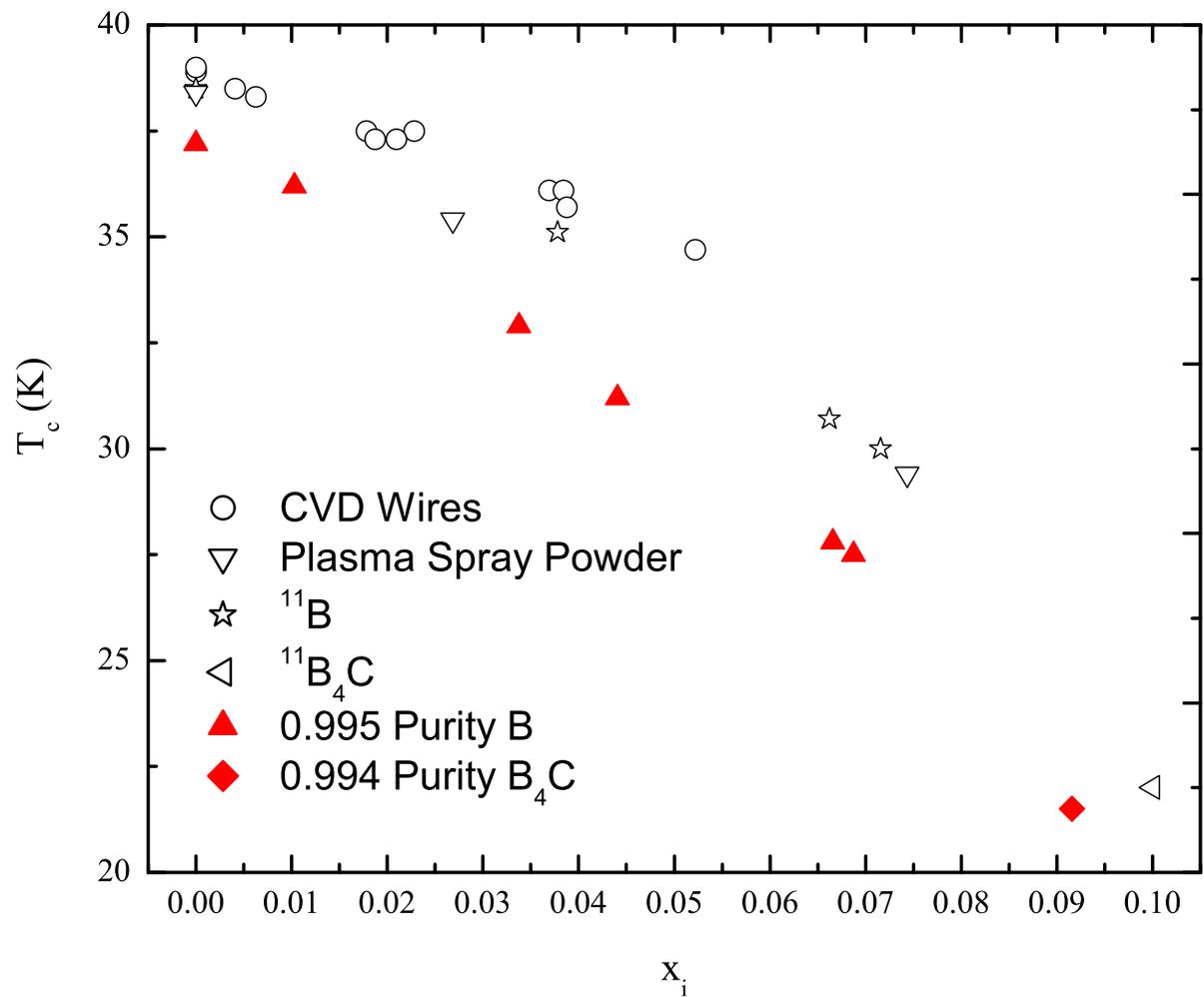}
\end{center}
\caption{Evolution of T$_c$ as a function of x$_i$. CVD wires and
powders are from references \cite{16} and \cite{20} respectively.
Data on the sample prepared with isotopically enriched
$^{11}$B$_4$C is from reference \cite{15}.}\label{f10}
\end{figure}

\clearpage

\begin{figure}
\begin{center}
\includegraphics[angle=0,width=180mm]{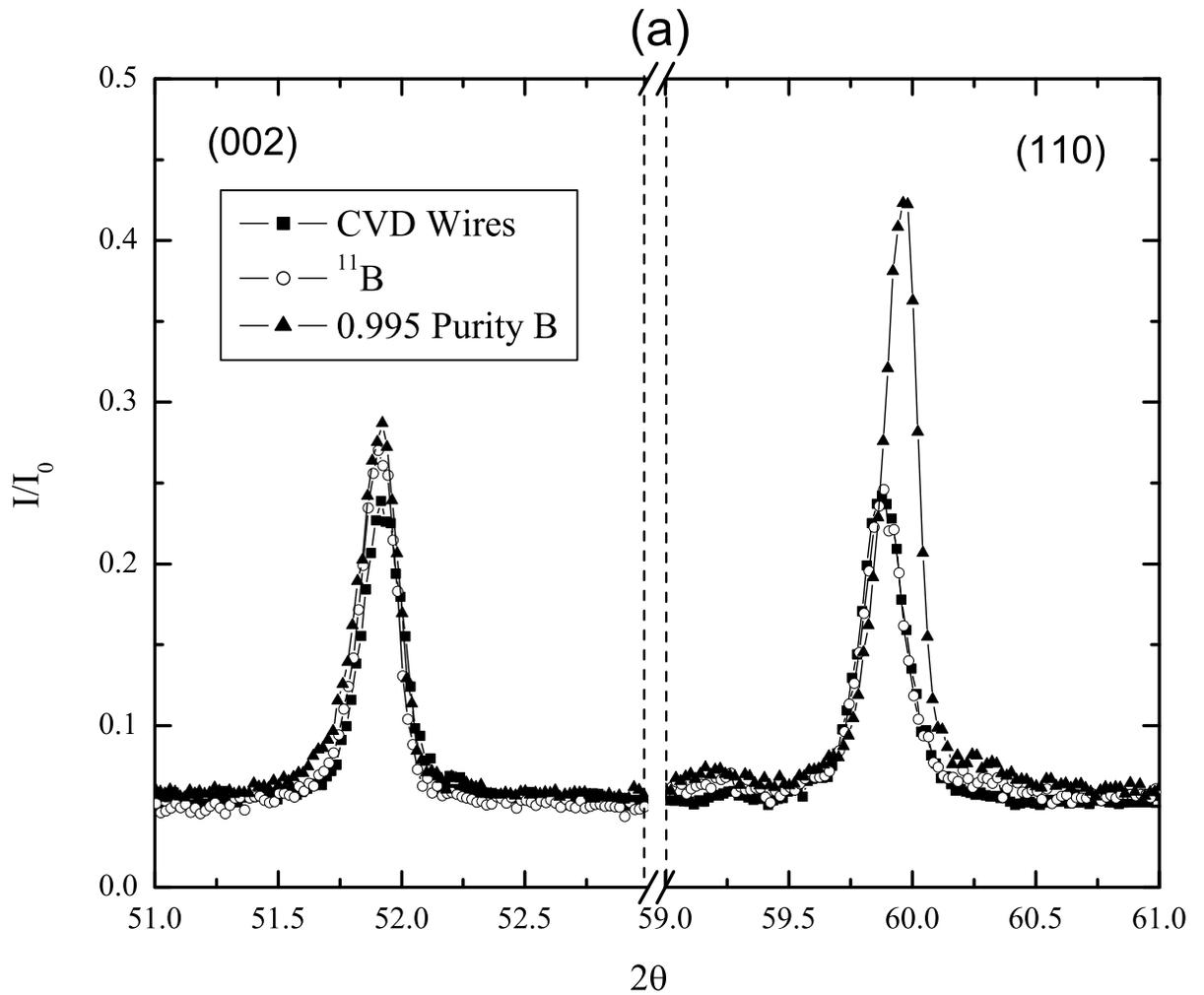}
\end{center}
\caption{Comparison of (002) and (110) x-ray diffraction peaks for
pure MgB$_2$ made using different purity boron as the starting
material. The 0.995 purity shows a shift in the (110) peak which
presumably is not a result of inadvertent carbon doping.
}\label{f11}
\end{figure}

\clearpage

\begin{figure}
\begin{center}
\includegraphics[angle=0,width=180mm]{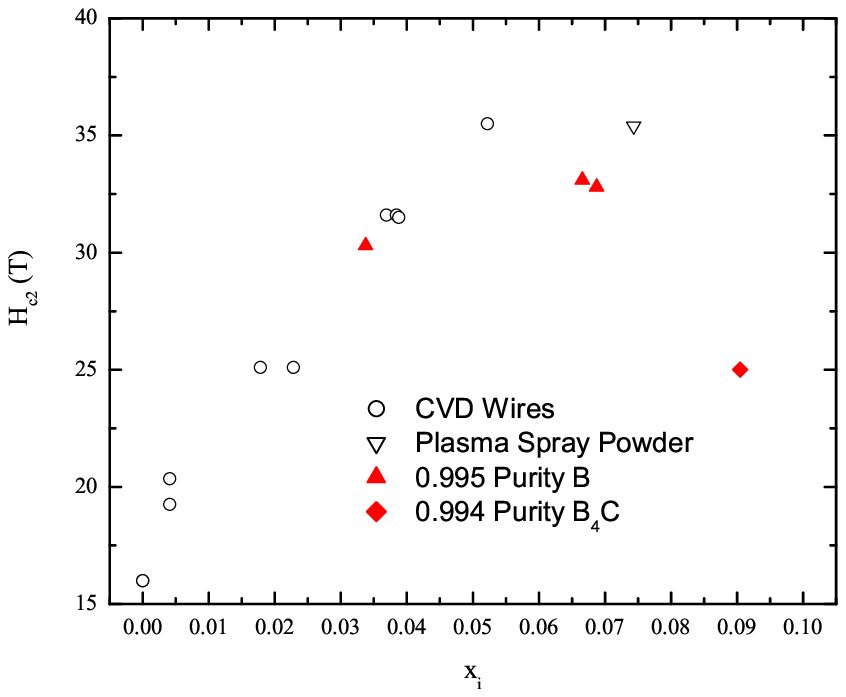}
\end{center}
\caption{H$_{c2}$(T=0) curves for the samples with different
purity in the starting boron. CVD wires and plasma spray powders
are from references \cite{16} and \cite{20} respectively. H$_{c2}$
data on the highest doping level is from reference
\cite{21}.}\label{f12}
\end{figure}


\begin{thebibliography}{00}
\bibitem {1} J. Nagamatsu, N. Nakagawa, T. Muranaka, Z. Takahiro, Y. Zenitani, and J. Akimitsu, "Superconductivity at 39K in Magnesium Diboride" Nature (London, United Kingdom) 410 (2001) 6824.
\bibitem {2} S.L. Bud'ko, C. Petrovic, G. Lapertot, C.E.
Cunningham, P.C. Canfield, M-H. Jung, and A.H. Lacerda, Phys. Rev.
B 63 (2001) 220503.
\bibitem {3} S.L. Bud'ko, V.G. Kogan, and P.C. Canfield, Phys.
Rev. B 64 (2001) 180506.
\bibitem {4} A.K. Pradham, Z.X. Shi, M. Tokunaga, T. Tamegai, Y.
Takano, K. Togano, H. Kito, and H. Ihara, Phys. Rev. B 64 (2001)
212509.
\bibitem {5} A.V. Sologubenko, J. Jun, S.M Kazakov, J. Karpinski,
and H.R. Ott, Phys. Rev. B 65 (2002) 180505.
\bibitem {6} M. Eisterer, M. Zehetmayer, and H.W. Weber, Phys.
Rev. Lett. 90 (2003) 247002.
\bibitem {7} R.H.T. Wilke, S.L. Bud'ko, P.C. Canfield, D.K. Finnemore, Raymond J. Suplinskas, and S.T. Hannahs, Phys. Rev. Lett. 91 (2004) 217003.
\bibitem {8} S.M. Kazakov, R. Puzniak, K. Rogacki, A.V. Mironov, N.D. Zhigadlo, J. Jun, Ch. Soltmann, B. Batlogg, and J. Karpinski, Phys. Rev.
B 71 (2005) 024533.
\bibitem {9} T. Masui, S. Lee, A. Yamamoto, K. Kajita, and S. Tajima, Physica C, Physica C 412-414 (2004) 303-306.
\bibitem {10} R. Puzniak, M. Angst, A. Szewczyk, J. Jun, S.M.
Kazakov, and J. Karpinski, cond-mat/0404579.
\bibitem {11} M. Angst, S.L. Bud'ko, R.H.T. Wilke, and P.C. Canfield, Phys. Rev. B 71 (2005) 144512.
\bibitem {16} R.H.T. Wilke,  S.L. Bud'ko, P.C. Canfield, D.K.
Finnemore, Raymond J. Suplinskas, and S.T. Hannahs, accepted for
publication in Physica C.
\bibitem {12} A. Serquis, L. Civale, X.Z. Liao, J.Y. Coulter, Y.T.
Zhu, M. Jaime, D.E. Peterson, F.M. Mueller, V.F. Nesterenko, and
Y. Gu, Appl. Phys. Lett. 82 (2003) 2847.
\bibitem {13} P. Lezza, V. Ab\"{a}cherli, N. Clayton, C. Senatore, D.
Uglietti, H.L. Suo, and R. Fl\"{u}kiger, Physica C 401 (2004) 305.
\bibitem {30} X.L. Wang, Q.W. Yao, J. Horvat, M.J. Qin, and S.X.
Dou, Supercond. Sci. Technol. 17 (2004) L21.
\bibitem {31} J. Wang, Y. Bugoslavsky, A. Ferenov, L. Cowey, A.D.
Caplin, L.F. Cohen, J.L. McManusDriscoll, L.D. Cooley, X. Song,
and D.C. Larbalestier, Appl. Phys. Lett. 81 (2002) 2026
\bibitem {32} Y. Zhao, Y. Feng, C.H. Cheng, L. Zhou, Y. Wu, T.
Machi, Y. Fudamoto, N. Koshizuka, and M. Murakami, Appl. Phys.
Lett. 79 (2001) 1154.
\bibitem {33} X.F. Rui, Y. Zhao, Y.Y. Xu, L. Zhang, S.F. Sun, Y.Z.
Wang, and H. Zhang, Supercond. Sci. Technol. 17 (2004) 689.
\bibitem {14} R.A. Ribeiro, S.L. Bud'ko, C. Petrovic, and P.C.
Canfield, Physica C 384 (2003) 227.
\bibitem {15} M. Avdeev, J.D. Jorgensen, R.A. Ribeiro, S.L. Bud'ko, P.C. Canfield, Physica C 387 (2003) 301.
\bibitem {23} S. Lee, Takahiko Masui, Ayako Yamamoto, Hiroshi Uchiyama, Setsuko Tajima, Physica C 397 (2003) 7.
\bibitem {19} R.A. Ribeiro, S.L. Bud'ko, C. Petrovic, and P.C.
Canfield, Physica C 382 (2002) 194.
\bibitem {77} P.C. Canfield, D.K. Finnemore, S.L. Bud'ko, J.E.
Ostensen, G. Lapertot, C.E. Cunningham, and C. Petrovic, Phys.
Rev. Lett. 86 (2001) 2423.
\bibitem {20} J.V. Marzik, R.J. Suplinskas, R.H.T. Wilke, P.C.
Canfield, D.K. Finnemore, M. Rindfleisch, J. Margolies, and S.T.
Hannahs, Physica C 423 (2005) 83.
\bibitem {21} Z. Ho\v{l}anov\'{a}, J. Ka\v{c}mar\v{c}\'{i}k,
P. Szab\'{o}, P. Samuely, I. Sheikin, R. A. Ribeiro, S. L. Bud'ko
and P. C. Canfield, Physica C 404 (2004) 195.
\bibitem {22} A. Gurevich, Phys. Rev. B 67 (2003) 184515.


\end{thebibliography}
\end{document}